# Stock Price Prediction Using Machine Learning and LSTM-Based Deep Learning Models

Sidra Mehtab[1], Jaydip Sen[2] and Abhishek Dutta[3]

Department of Data Science and Artificial Intelligence, Praxis Business School.
Bakrahat Road, P.O. Rasapunja, Off Thakurpukur Road, Kolkata 700104, INDIA
{[1]smehtab@acm.org, [2]jaydip@praxis.ac.in, [3]duttaabhishek0601@gmail.com}

**Abstract.** Prediction of stock prices has been an important area of research for a long time. While supporters of the *efficient market hypothesis* believe that it is impossible to predict stock prices accurately, there are formal propositions demonstrating that accurate modeling and designing of appropriate variables may lead to models using which stock prices and stock price movement patterns can be very accurately predicted. Researchers have also worked on technical analysis of stocks with a goal of identifying patterns in the stock price movements using advanced data mining techniques. In this work, we propose an approach of hybrid modeling for stock price prediction building different machine learning and deep learning-based models. For the purpose of our study, we have used NIFTY 50 index values of the National Stock Exchange (NSE) of India, during the period December 29, 2014 till July 31, 2020. We have built eight regression models using the training data that consisted of NIFTY 50 index records during December 29, 2014 till December 28, 2018. Using these regression models, we predicted the *open* values of NIFTY 50 for the period December 31, 2018 till July 31, 2020. We, then, augment the predictive power of our forecasting framework by building four deep learning-based regression models using long-and short-term memory (LSTM) networks with a novel approach of walk-forward validation. Using the grid-searching technique, the hyperparameters of the LSTM models are optimized so that it is ensured that validation losses stabilize with the increasing number of epochs, and the convergence of the validation accuracy is achieved. We exploit the power of LSTM regression models in forecasting the future NIFTY 50 *open* values using four different models that differ in their architecture and in the structure of their input data. Extensive results are presented on various metrics for all the regression models. The results clearly indicate that the LSTM-based univariate model that uses one-week prior data as input for predicting the next week's *open* value of the NIFTY 50 time series is the most accurate model.

## 1 Introduction

Prediction of future movement of stock prices has been an area that attracted the attention of the researchers over a long period of time. While those who support the school of thought of the *efficient market hypothesis* believe that it is impossible to predict stock prices accurately, there are formal propositions demonstrating that with the choice of appropriate variable and suitable modeling, it is possible to predict the future stock prices and stock price movement patterns, with a fairly high level of accuracy. In this regard, Sen and Datta Chaudhuri demonstrated a new approach to stock price prediction using the decomposition of time series [1-8]. In addition, a granular approach of stock price prediction in a short-term forecast horizon has been proposed by Sen that uses powerful capabilities of machine learning and deep learning models [9-10].

Mehtab and Sen present a highly robust and reliable predictive framework for stock price prediction by combining the power of text mining and natural language processing in machine learning models like regression and classification [11]. By analyzing the sentiments in the social media and utilizing the sentiment-related information in a non-linear multivariate regression model based on *self-organizing fuzzy neural networks* (SOFNN), the authors have demonstrated a high level of accuracy in predicted values of NIFTY index values. In another recent work, Mehtab and Sen presented a suite of *convolutional neural network* (CNN)-based models, for achieving a high level of accuracy and robustness in forecasting on a multivariate financial time series data [28, 29].

Researchers have proposed models on technical analysis of stock prices wherein the goal is to detect patterns in stock movements that lead to profit for the investors. For this purpose, various economic and stock price-related indicators have been proposed in the literature. Some of these indicators are: Bollinger Band, *moving average convergence divergence* (MACD), *relative strength index* (RSI), *moving average* (MA), *momentum stochastics* (MS), *meta sine wave* (MSW). In addition to these indicators, some of the well-known patterns in stock price movements like *head and shoulders*, *triangle*, *flag*, *Fibonacci fan*, *Andrew's pitchfork,* etc., are also considered as important indicators for investment in the stock market. These approaches provide effective visualizations to potential investors in making the right investment decisions.

The current work proposes a gamut of machine learning and deep learning-based predictive models for accurately predicting the NIFTY 50 stock price movement in NSE of India. The historical index values of NIFTY 50 for the period December 29, 2014 till December 28, 2018 has been used as the training dataset. Using the training dataset, the predictive models are built, and using the models, the *open* values of the NIFTY 50 index are predicted for the test period that spanned over the time horizon December 31, 2018 till July 31, 2020. The predictive power of the models is further enhanced by introducing the powerful deep learning-based *long- and short-term memory* (LSTM) network into the predictive framework. Four LSTM models have been built in this work. The models have different architectures and different structures in their input data. While three LSTM models are based on univariate data, one model is a multivariate one. From the input data point of view, three models used the previous two weeks' data as their input for forecasting the *open* values of the NIFTY 50 time series for the next week, while one model used only one-week prior data as the input.

The rest of the paper is organized as follows. In Section 2, we explicitly define the problem at hand. Section 3 provides a brief review of the related work on stock price movement prediction. In Section 4, we describe our research methodology. Extensive results on the performance of the predictive models are presented in Section 5. This section describes the details of all the predictive models that are built in this work and the results they have produced. Finally, Section 6 concludes the paper.

## 2   Problem Statement

The goal of our work is to collect the stock price of NIFTY 50 from the NSE of India over a reasonably long period of five and half years and develop a robust forecasting framework for forecasting the NIFTY 50 index values. We hypothesize that it is possible for a machine learning or a deep learning model to learn from the features of the past movement patterns of daily NIFTY 50 index values, and these learned features can be effectively exploited in accurately forecasting the future index values of the NIFTY 50 series. In the current proposition, we have chosen a forecast horizon of one year for the machine learning models, and one week for the deep learning models and demonstrated that the future NIFTY index values can be predicted using these models with a fairly high level of accuracy. To validate our hypothesis, in our past work, we used CNN-based deep learning models to build highly accurate predictive frameworks for forecasting future NIFTY 50 index values [28]. In the present work, we follow four different approaches in building *long and short-term memory* (LSTM) network-based models in order to augment the predictive power of our forecasting models. It must be noted that in this work, we are not addressing the issues of short-term forecasting which are of interest to the intra-day traders. Instead, the propositions in this paper are relevant for medium-term investors who might be interested in a weekly forecast of the NIFTY 50 index values.

## 3   Related Work

The currently existing work in the literature on time series forecasting and stock price prediction can be broadly categorized in three clusters, based on the use of variables and the approach to modeling the problem. The first category of work mainly consists of models that use bivariate or multivariate regression on cross-sectional data [12-16]. Due to their inherent simplicity and invalidity of the linearity assumptions that they make, these models fail to produce highly accurate results most of the time. The propositions in the second category utilize the concepts of time series and other econometric techniques like *autoregressive integrated moving average* (ARIMA), Granger Causality Test, *autoregressive distributed lag* (ARDL), *vector autoregression* (VAR), and *quantile regression* to forecast stock prices [17-20]. The third category of work includes learning-based approaches propositions using machine learning, deep learning, and natural language processing [21-24].

   Except for the category of work that utilizes learning-based approaches, one of the major shortcomings of the current propositions in literature for stock price prediction is their inability to accurately predict highly dynamic and fast-changing patterns in stock price movement. In this work, we attempt to address the problem by exploiting the power of machine learning and deep learning-based models in building a very robust, reliable, and accurate framework for stock index prediction. In particular, we have used a *long-and-short-term memory* (LSTM) network-based deep learning model and studied its performance in predicting future stock index values.

## 4 Methodology

In Section 2, we mentioned that the goal of this work is to develop a predictive framework for forecasting the daily price movement of NIFTY 50. We collect the historical index values of NIFTY 50 for the period: December 29, 2014 till July 31, 2020 from the Yahoo Finance website [25]. The raw NIFTY 50 index values consist of the following variables: (i) *date*, (ii) *open* value of the index, (iii) *high* value of the index, (iv) *low* value of the index, (v) *close* value of the index, and (vi) *volume* of the stock traded on a given date.

We followed the approach of regression in forecasting the NIFTY 50 index values. For this purpose, we used the variable *open* as the response variable and the other variables as the predictors. We carried out some pre-processing of the data before using it in training and testing the regression models. We design the following derived variables using the *six* variables in the raw NIFTY 50 index records. These derived variables will be used for building predictive models.

The following five variables are derived and used in our forecasting models:

a) *high_norm*: it refers to the normalized values of the variable *high*. We use *min-max normalization* to normalize the values. Thus, if the maximum and the minimum values of the variable *high* are $H_{max}$ and $H_{min}$ respectively, then the normalized value *high_norm* is computed as: *high_norm* = (*high* - $H_{min}$)/($H_{max}$ – $H_{min}$). After the normalization operation, all values of *high_norm* lie inside the interval [0, 1].

b) *low_norm*: this normalized variable is computed from the variable *low* in a similar way as *high_norm* is computed: *low_norm* = (*low* – $L_{min}$)/($L_{max}$ - $L_{min}$). The values of *low_norm* also lie in the interval [0, 1].

c) *close_norm*: it is the normalized version of the variable *close*, and is computed as: *close_norm* = (*close* - $C_{min}$) / ($C_{max}$ – $C_{min}$). The interval in which the values of this variable lie is [0, 1].

d) *volume_norm*: this variable is the normalized value of the variable *volume*. It is computed in a similar way as *high_norm*, *low_norm*, and the *close_norm*, and its values also lie in the interval [0, 1].

e) *range_norm*: this variable is the normalized counterpart of the variable *range*. The *range* for a given index record is computed as the difference between the *high* and the *low* values for that index record. Like all other normalized variables e.g., *high_norm*, *low_norm*, or *close_norm*, the variable *range_norm* also lies in the closed interval [0, 1].

After we carry out the pre-processing and transformation of the variables on the NIFTY 50 data for the period December 29, 2014–July 31, 2020, we use the processed data for building and testing the regression models based on machine learning and deep learning.

For training the regression models, we use the data for the period December 29, 2014 (which was a Monday) till December 28, 2018 (which was a Friday). The models are then tested on the data for the period December 31, 2018 – a Monday - till July 31, 2020 – a Friday. The data is collected from the Yahoo Finance website and these are daily NIFTY 50 index values. The training dataset consisted of 1045 records that included NIFTY 50 index data for 209 weeks. On the other hand, there were 415 records in the test dataset encompassing 83 weeks. For the machine learning-based models, we used the daily data in the training set to construct the models, and then we predicted the *open* values of the NIFTY 50 index for every day in the test dataset. For building the deep learning-based LSTM models, however, we follow a different approach. The approach is called *multi-step forecasting with walk-forward validation* [27]. Following this approach, we build the models using the records in the training dataset and then deploy the model for forecasting the *open* value of the NIFTY 50 index on a weekly basis for the records in the test dataset. As soon as the week for which the last round of forecasting was made was over, the actual records for that week were included in the training dataset for the purpose of forecasting the next week's *open* values of the NIFTY 50 index. As a working week in the NSE involves five days - Monday through Friday – each round of forecasting resulted in five values corresponding to the predicted *open* values for the five days in the upcoming week.

For building the machine learning-based regression models, we considered two cases, which we discuss below.

**Case I:** As already been mentioned earlier, the training dataset included historical records of NIFTY 50 index values for the period December 29, 2014 till December 28, 2018. The training dataset included index values for 1045 days. In *Case I*, the performance of the models was tested in terms of the accuracy with which they could predict the *open* values for NIFTY 50 index records of the training dataset. In other words, in *Case I*, we evaluate the training performance of the machine learning-based regression models. The predictions are made on daily basis.

**Case II:** In this case, the predictive models are tested on the test dataset and their performance is evaluated. The test data consists of historical records of NIFTY 50 index values for the period December 31, 2018 till July 31, 2020. The performances of the models are evaluated in terms of their prediction accuracy of *open* values for each of the 415 days included in the test dataset. Hence, in *Case II*, we have evaluated the test performance of the machine learning models.

In this work, we designed and evaluated eight machine learning-based regression models. These models are: (i) *multivariate linear regression*, (ii) *multivariate adaptive regression spline* (MARS), (iii) *regression tree*, (iv) *bootstrap aggregation* (Bagging), (v) *extreme gradient boosting* (XGBoost), (vi) *random forest* (RF), (vii) *artificial neural network* (ANN), and (viii) *support vector machine* (SVM). For the purpose of evaluation of performances of these model, we use two metrics. The first metric that we use for evaluating a regression model is the value of the *product-moment correlation coefficient* between the *actual* and the *predicted* values of the *open* values of the NIFTY 50 index. The models exhibiting higher values of correlation coefficient are supposed to be more accurate. The second metric that we use for model evaluation is the ratio of the *root mean square error* (RMSE) values to the mean of the actual *open* values in the dataset. The models that yield lower values of this ratio are more accurate.

To make our forecasting framework more robust and accurate, we build some deep learning-based regression models too. In one of our previous work, we demonstrated the efficacy and effectiveness of *convolutional neural networks* (CNNs) in forecasting time series index values [28]. In this work, we have utilized the predictive power of another deep learning model – *long- and short-term memory* (LSTM) networks - in forecasting on a complex multivariate time series like the NIFTY 50 series. LSTM is a special type of *recurrent neural networks* (RNNs) – neural networks that allow feedback loops to communicate data from a node in a forward layer to a node in a backward layer [26]. In RNN networks, the output of the network at a given time slot is dependent on the input to the network in the given time slot along with the state of the network in the previous time slot. However, RNNs suffer from a problem known as *vanishing and exploding gradient problem*, in which a network either stops learning or continues to learn at a very high learning rate so that it never converges to the point of the minimum error [26]. LSTM networks overcome the problem of vanishing and exploding gradient problems by intelligently forgetting some past irrelevant information, and hence such network proves very suitable for modeling sequential data, like texts and time series. LSTM networks consist of memory cells that maintain their state information over time using memory and gating units that regulate and control information flow through them. Three types of gates are used in an LSTM network – *forget gates*, *input gates*, and the *output gates*. The forget gates are instrumental in throwing away irrelevant past information, and in remembering only that information which is relevant at the current slot. The input gates control the new information that acts as the input to the current state of the network. The old information from the forget gates and the new information from the input gates are effectively aggregated by the cell state vector. Finally, the output gates produce the output from the network at the current slot. This output can be considered as the forecasted value computed by the model for the current slot. The architecture of LSTM networks integrated with the *backpropagation through time* (BPTT) algorithm for learning the parameters provides these networks with a high degree of power in forecasting in univariate and multivariate time series [26].

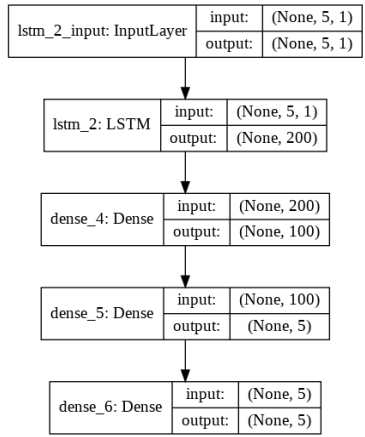

**Fig. 1.** The architecture of univariate LSTM model with prior one week's data as the input

We exploit the power of LSTM models in multi-step time series forecasting using a *walk-forward validation method* [27]. In this method, a model is required to make a one-week prediction, and the actual data for that week is used in the model for making the forecast for the next week. This is both realistic and practical, as in most of the real-world applications, forecast horizon longer than one week is not used.

We have used four different LSTM models in this work. The approaches vary in architectures of the models and also on the shape of the input data the models use. The four models are: (i) *LSTM model for multi-step forecasting with univariate input data of one week*, (ii) *LSTM model for multi-step forecasting with univariate input data of two weeks*, (iii) *Encoder-decoder LSTM for multi-step forecasting with univariate input data for two weeks*, and (iv) *Encoder-decoder LSTM for multi-step forecasting with multivariate input data for two weeks*.

The architectural design and the parameters of each of the four models are now discussed in the following.

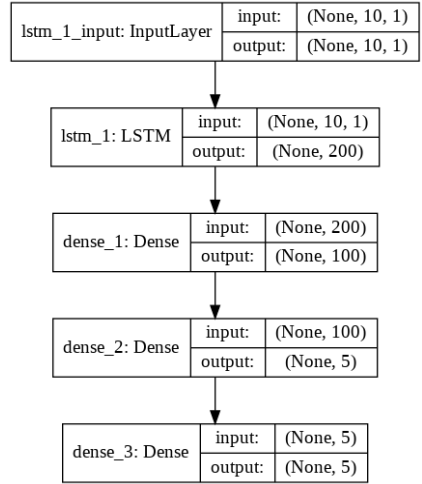

**Fig. 2.** The architecture of univariate LSTM model with prior two week's data as the input

The first model - univariate LSTM model with prior one week's data as the input - performs multi-step time series forecasting using only the univariate sequence of the *open* values of the NIFTY 50 time series. We train the model using the training dataset records, and then use the model to forecast the *open* values for the next week (i.e., the next five values as a week consists of five working days). The forecasting is being done in a multi-step manner with a walk-forward validation mode. The details of the design of each layer and the overall architecture of the model are as follows. The shape of the input data to the input layer of the network is (5,1) indicating that the previous five values (i.e., one week's data) of the time series are used as the input, and only one attribute of the data (i.e., the *open* value) is considered. The input layer passes the data onto the LSTM layer that has 200 nodes at the output with the ReLU activation function being used in those nodes. The output of the LSTM layer is passed onto a dense layer that has 200 nodes at its input, and 100 nodes with ReLU activation function at the output. The dense layer uses *mean square error* (MSE) as the *loss function* and ADAM as the *optimizer*. The dense layer is finally connected to the output layer that is also a fully-connected layer. The output layer of the model has 100 nodes at its input and 5 nodes at the output. The 5 nodes at the output produce the forecasted values for the five days of the next week. Again, the nodes at the output layer use MSE as the *loss function* and ADAM as the *optimizer*. Fig. 1 depicts the architecture of the first LSTM model, which we will refer to as LSTM#1.

The second LSTM model, which we refer to as LSTM#2, is also a univariate model that uses the previous two weeks' *open* values as the input and yields the forecast for the next five days (i.e., for the next week). The architecture and other parameters of the model remain identical to those of the first model (i.e., LSTM#1). The only change that is introduced is that the input to the model, in this case, is the previous two week's *open* values. Fig. 2 depicts the architecture of the model.

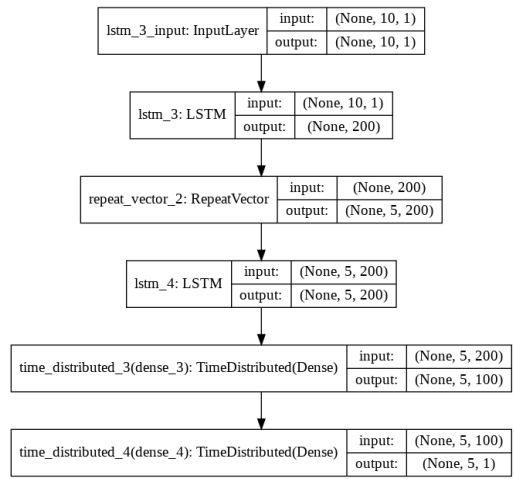

**Fig. 3.** The architecture of univariate encoder-decoder LSTM model with prior two weeks' data as the input

The third LSTM model – encoder-decoder LSTM model with univariate data of the previous two weeks as the input - does not produce a vector sequence as its output directly, unlike the previous two models. In fact, the model consists of two sub-models: the encoder sub-model reads and encodes the input sequence, while the decoder sub-model reads the encoded input sequence, and makes a one-step prediction for each element in the output sequence. We have employed LSTM in the decoder sub-module of the model that enables the model to be aware of the values that were predicted for the prior day in the predicted output sequence and utilize that information in the prediction of its next value. Fig. 3 shows the architecture of the model.

The first LSTM layer consists of 200 nodes at its output with each node having a ReLU activation function. This layer acts as the decoder sub-module that reads the input sequence having the shape (10, 1). The shape of the input data indicates that the time series is univariate with only the *open* value being considered for the previous two weeks' data as input. The LSTM layer produces a 200-element vector (one output per node) that captures deep features from the input sequence of 10 values. For each time-step in the output sequence that the model produces, the internal representation of the input sequence is repeated multiple times, once for each output sequence. The data shape form output of the repeat vector layer is (5, 200) that corresponds to the five time-stamps in the output sequence, and the 200 features being extracted by the LSTM layer working as a decoder. An additional second LSTM decoder layer performs decoding of the output sequence using its 200 units (i.e., nodes). Essentially, each of the 200 nodes will yield a value for each of the five days in a week. This represents a basis for the predicted value for each day in the output sequence. The output sequence of the second LSTM decoder is passed through a fully-connected layer that interprets each value in the output sequence before it is sent to the final output layer. Finally, the output layer produces the prediction for a single step (i.e., for a single day) at each step, not for all the five steps in a single round. The same fully connected layer and output layer are used to process each time-step provided by the decoder LSTM. This is achieved by using a *TimeDistributed wrapper* that packs the interpretation layer and the output layer in a time-synchronized manner allowing the use of wrapped layers in an identical manner for each time-step from the decoder. This feature enables the decoder LSTM and the wrapped dense layers in understanding the context of each step in the output sequence while reusing the same weights to perform the interpretation. The output of the model, in this case, is a three-dimensional vector with the same structure as the input – each output consisting of [*samples*, *timestamps*, *features*]. We have a single feature – the *open* value of the NIFTY 50 index. A single-week prediction will, therefore, have the shape [None, 5, 1]. The structure of the output of this model is thus different from the first two LSTM models, both of which were of the shape [None, 5]. While we used the ReLU activation function in the output of the two decoder LSTM layers and the *TimeDistributed Dense layer*, at the final output layer of the model, MSE and ADAM were used as the *loss function* and the *optimizer* respectively. We refer to this model as LSTM#3.

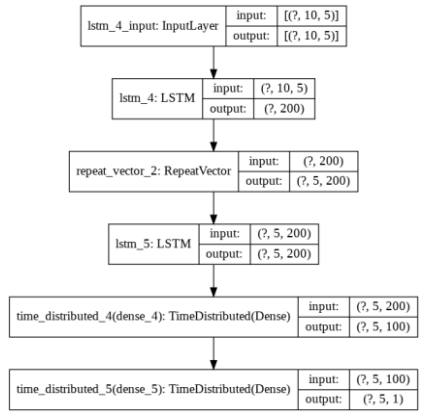

**Fig. 4.** The architecture of multivariate encoder-decoder LSTM model with prior two week's data as the input

The fourth and the last model of our current work is based on an encoder-decoder LSTM that uses multivariate input. In other words, instead of using a single input variable *open* as the input as it was done by the three models previously discussed, this model uses all the five variables – *open*, *high*, *low*, *close*, and *volume* – to forecast the value of *open*. The model is built by using each one-dimensional time series corresponding to each of the input variables as a separate sequence of input. The LSTM creates an internal representation of each input sequence and combines them together before interpreting and decoding the combined representation. The model is the most complex model among all the models that we have proposed in this work.

**Table 1.** Multivariate regression results

| Stock | Case I Training Data | | Case II Test Data | |
|---|---|---|---|---|
| NIFTY 50 | Correlation | 0.99 | Correlation | 0.99 |
| | RMSE | 0.27 | RMSE | 0.42 |

**Table 2.** MARS regression results

| Stock | Case I Training Data | | Case II Test Data | |
|---|---|---|---|---|
| NIFTY 50 | Correlation | 0.99 | Correlation | 0.99 |
| | RMSE | 0.42 | RMSE | 0.85 |

**Table 3.** Decision tree regression results

| Stock | Case I Training Data | | Case II Test Data | |
|---|---|---|---|---|
| NIFTY 50 | Correlation | 0.98 | Correlation | 0.16 |
| | RMSE | 2.52 | RMSE | 10.40 |

**Table 4.** Bagging regression results

| Stock | Case I Training Data | | Case II Test Data | |
|---|---|---|---|---|
| NIFTY 50 | Correlation | 0.99 | Correlation | 0.96 |
| | RMSE | 1.75 | RMSE | 3.72 |

## 5   Performance Results

This Section provides a detailed discussion on the performance results of all the predictive models that we have constructed and tested in this work. As mentioned in Section 4, we designed two metrics for evaluating the performance of the machine learning-based regression models. These metrics are: (i) product-moment correlation coefficient between the actual and the predicted *open* NIFTY 50 index values, and (ii) the ratio of the RMSE and the mean of the actual *open* NIFTY 50 index values in the dataset. In Tables 1 – 8, we have presented the performance results of the machine learning-based regression models. The performances of all models in training and tests are presented. Since the test performance is the one that matters, we observe that multivariate regression, MARS, and random forest have outperformed all other models on the metric correlation coefficient among the actual and the predicted *open* values in the test dataset. However, the lowest ratio of RMSE to the mean of the actual *open* values was yielded by the multivariate regression and the random forest. Hence, on the basis of the performances of all machine learning models, we conclude that the multivariate regression and the random forest regression were the most accurate models in terms of their forecasting accuracies on the NIFTY 50 time series.

Table 5. Boosting regression results

| Stock | Case I Training Data | | Case II Test Data | |
|---|---|---|---|---|
| NIFTY 50 | Correlation | 0.99 | Correlation | 0.98 |
| | RMSE | 0.37 | RMSE | 1.87 |

Table 6. Random forest regression results

| Stock | Case I Training Data | | Case II Test Data | |
|---|---|---|---|---|
| NIFTY 50 | Correlation | 0.99 | Correlation | 0.99 |
| | RMSE | 0.29 | RMSE | 0.42 |

Table 7. ANN regression results

| Stock | Case I Training Data | | Case II Test Data | |
|---|---|---|---|---|
| NIFTY 50 | Correlation | 0.67 | Correlation | 0.44 |
| | RMSE | 12.77 | RMSE | 19.31 |

Table 8. SVM regression results

| Stock | Case I Training Data | | Case II Test Data | |
|---|---|---|---|---|
| NIFTY 50 | Correlation | 0.99 | Correlation | 0.58 |
| | RMSE | 0.75 | RMSE | 8.40 |

We now present the performance results of the deep learning models. The details of the design of all the four models were presented in Section 4. The performance of each of the models is evaluated by executing it on the test data over 10 rounds. For each round, we have observed its overall RMSE value of a week, the RMSE values for the individual days in a week (i.e., Monday – Friday), the time the model took for completing its execution, and the ratio of the RMSE to the mean of the actual *open* value in the test dataset. It may be noted here that the number of records in the training and the test dataset was 1045 and 415 respectively. The mean *open* value in the test dataset was 11070.59.

Table 9 presents the performance results of the LSTM#1 model. The model was trained on a hardware system consisting of an Intel i5-8250U processor with clock speed 1.60 GHz – 1.80 GHz, 8 GB RAM, and running 64-bit Windows 10 operating system. The unit of measurement for all execution time-related records is *second*. We observe that the LSTM#1 model took 18.64s on an average, and it yielded a mean value of the ratio of RMSE to the mean of the *open* values in the test dataset as 0.0311. It is also interesting to note that the mean RMSE values consistently increased from Monday through Friday. Fig. 5 presents the performance results of the LSTM#1 model for round #2 presented in Table 9.

**Table 9.** LSTM regression results – univariate time series with previous week data as the training input (LSTM#1)

| No. | RMSE | Mon | Tue | Wed | Thu | Fri | Time |
|---|---|---|---|---|---|---|---|
| 1 | 350.7 | 250 | 295 | 343 | 396 | 437 | 19.14 |
| 2 | 347.2 | 232 | 303 | 341 | 390 | 435 | 16.42 |
| 3 | 351.9 | 243 | 296 | 349 | 398 | 439 | 19.80 |
| 4 | 323.6 | 210 | 273 | 305 | 369 | 419 | 19.36 |
| 5 | 347.4 | 253 | 285 | 336 | 388 | 441 | 18.52 |
| 6 | 314.5 | 201 | 259 | 299 | 359 | 411 | 18.54 |
| 7 | 330.8 | 234 | 276 | 322 | 369 | 419 | 18.85 |
| 8 | 340.1 | 228 | 278 | 326 | 393 | 434 | 18.55 |
| 9 | 378.1 | 251 | 385 | 341 | 412 | 467 | 18.87 |
| 10 | 361.5 | 219 | 284 | 338 | 450 | 456 | 18.35 |
| **Mean** | **344.57** | 232 | 293 | 330 | 392 | 436 | **18.64** |
| **Min** | 314.5 | 201 | 259 | 299 | 359 | 411 | 16.42 |
| **Max** | 378.1 | 253 | 385 | 349 | 450 | 467 | 19.80 |
| **SD** | 18.47 | 17.9 | 34.5 | 16.7 | 25.8 | 17.0 | 0.899 |
| **RMSE/Mean** | **0.0311** | 0.02 | 0.03 | 0.03 | 0.04 | 0.04 | |

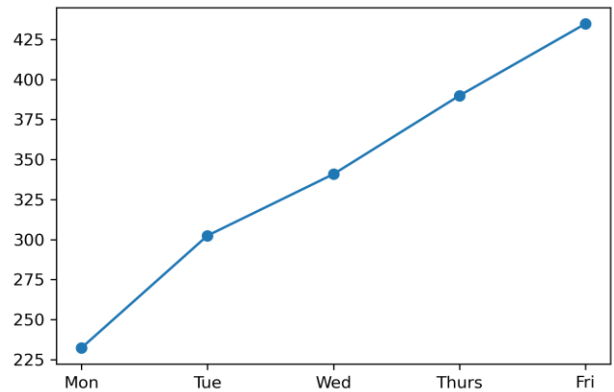

**Fig. 5.** Day-wise RMSE of LSTM#1 – univariate time series with one-week data as the input

Table 10 presents the performance of the deep learning regression model LSTM#2. The mean execution time for the 10 rounds of execution of the model on the same computing environment was found to be 31.44s. This was almost two times the time needed for the execution of the LSTM#1 model. The average value the ratio of the

RMSE to the mean of the actual *open* values yielded by the model was 0.0353, while the mean RMSE was 390.46. Hence, in terms of both the metrics – RMSE to mean *open* value and the mean execution time – the LSTM#2 model is found to be inferior to the LSTM#1 model.

**Table 10.** LSTM regression results – univariate time series with the previous two weeks' data as the training input (LSTM#2)

| No. | RMSE | Mon | Tue | Wed | Thu | Fri | Time |
|---|---|---|---|---|---|---|---|
| 1 | 393.1 | 336 | 363 | 390 | 439 | 429 | 31.29 |
| 2 | 369.4 | 293 | 343 | 373 | 413 | 411 | 31.40 |
| 3 | 368.0 | 318 | 346 | 363 | 410 | 396 | 31.36 |
| 4 | 431.9 | 367 | 409 | 398 | 528 | 440 | 31.95 |
| 5 | 397.9 | 343 | 383 | 376 | 452 | 425 | 31.49 |
| 6 | 391.6 | 318 | 360 | 397 | 439 | 429 | 31.45 |
| 7 | 408.2 | 356 | 414 | 397 | 448 | 421 | 31.44 |
| 8 | 363.9 | 304 | 337 | 357 | 406 | 405 | 31.53 |
| 9 | 395.1 | 345 | 369 | 404 | 438 | 413 | 31.03 |
| 10 | 385.5 | 322 | 353 | 389 | 435 | 418 | 31.44 |
| **Mean** | **390.46** | 330 | 367 | 385 | 441 | 419 | **31.44** |
| **Min** | 363.9 | 293 | 337 | 357 | 406 | 396 | 31.03 |
| **Max** | 432 | 367 | 414 | 404 | 528 | 440 | 31.95 |
| **SD** | 20.5 | 23.2 | 26.6 | 16.4 | 34.7 | 12.9 | 0.23 |
| **RMSE/Mean** | **0.0353** | 0.03 | 0.03 | 0.03 | 0.04 | 0.04 | |

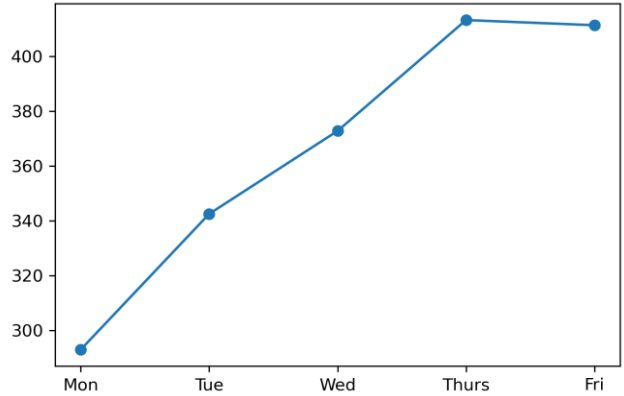

**Fig. 6.** Day-wise RMSE of LSTM#2 – univariate time series with previous two weeks' data as the input

Unlike the LSTM#1 model, model #LSTM#2 exhibited a different behavior in its RMSE. While the RMSE values for the model LSTM#2 increased consistently from Monday till Thursday in a week, the RMSE values for Friday were found to be smaller than those for Thursday. Fig. 6 presents the performance results of the LSTM#2 model for round #2 presented in Table 10.

Table 11 presents the performance results for the model LSTM#3 – the univariate encoder-decoder LSTM model. The mean time for execution of the model for 10 rounds of execution was found to be 14.53s, while the ratio of mean RMSE to the mean of the *open* values was 0.0369. Thus, while the model LSTM#3 is found to be marginally faster in execution when compared to the model LSTM#1, the latter is more accurate in its forecasting performance. The model LSTM#3 exhibited similar behavior in weekly RMSE values as the model LSTM#2. RMSE values increased from Monday till Thursday before experiencing a fall on Friday. Fig. 7 presents the performance results of the model LSTM#3 for round #9 presented in Table 11.

**Table 11.** Encoder decoder LSTM regression results – univariate time series with previous two weeks' data as the training input (LSTM#3)

| No. | RMSE | Mon | Tue | Wed | Thu | Fri | Time |
|---|---|---|---|---|---|---|---|
| 1 | 391.7 | 318 | 383 | 395 | 433 | 418 | 12.79 |
| 2 | 418.1 | 367 | 398 | 415 | 459 | 446 | 12.56 |
| 3 | 409.1 | 334 | 381 | 403 | 462 | 452 | 14.95 |
| 4 | 423.0 | 365 | 400 | 413 | 467 | 461 | 14.74 |
| 5 | 403.4 | 326 | 414 | 389 | 424 | 453 | 14.79 |
| 6 | 397.9 | 349 | 379 | 393 | 440 | 422 | 14.68 |
| 7 | 389.8 | 344 | 384 | 372 | 425 | 418 | 15.11 |
| 8 | 395.6 | 327 | 362 | 391 | 445 | 440 | 15.44 |
| 9 | 449.0 | 343 | 387 | 468 | 527 | 493 | 14.95 |
| 10 | 412.1 | 348 | 382 | 411 | 456 | 453 | 15.26 |
| **Mean** | **408.97** | 342 | 387 | 405 | 454 | 446 | **14.53** |
| **Min** | 389.8 | 318 | 362 | 372 | 424 | 418 | 12.56 |
| **Max** | 449.0 | 367 | 414 | 468 | 527 | 493 | 15.44 |
| **SD** | 17.92 | 16.2 | 14.0 | 25.7 | 29.3 | 22.9 | 1.00 |
| **RMSE/Mean** | **0.0369** | 0.03 | 0.04 | 0.04 | 0.04 | 0.04 | |

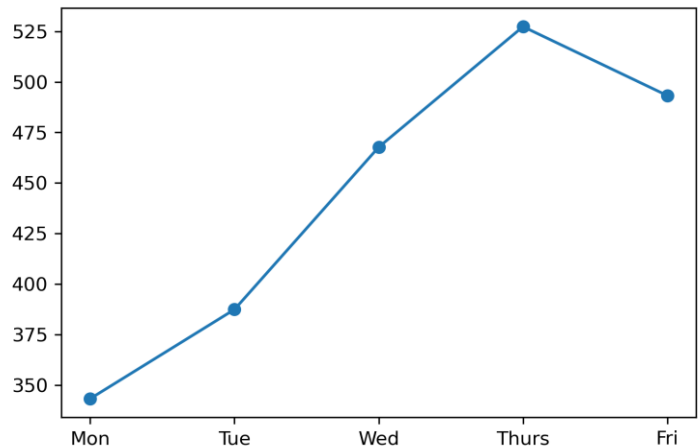

**Fig. 7.** Day-wise RMSE of LSTM#3 – univariate encoder-decoder with previous two weeks' data as the input

**Table 12.** Encoder decoder LSTM regression results – multivariate time series with previous two weeks' data as the training input (LSTM#3)

| No. | RMSE | Mon | Tue | Wed | Thu | Fri | Time |
|---|---|---|---|---|---|---|---|
| 1 | 1329.8 | 1396 | 1165 | 949 | 1376 | 1655 | 72.79 |
| 2 | 2798.8 | 3253 | 3131 | 2712 | 2407 | 2369 | 69.01 |
| 3 | 2764.2 | 2588 | 2926 | 2761 | 3100 | 2391 | 79.88 |
| 4 | 1754.2 | 1402 | 1856 | 1766 | 1543 | 2114 | 62.76 |
| 5 | 1217.4 | 1521 | 1098 | 1045 | 1001 | 1340 | 59.61 |
| 6 | 1162.9 | 1421 | 1100 | 1075 | 920 | 1239 | 72.28 |
| 7 | 2485.4 | 2034 | 2108 | 2258 | 2767 | 3091 | 68.76 |
| 8 | 1788.6 | 1280 | 1590 | 1705 | 1962 | 2252 | 62.82 |
| 9 | 1451.6 | 1921 | 1317 | 1367 | 1362 | 1179 | 62.42 |
| 10 | 2185.6 | 1191 | 1373 | 1901 | 2601 | 3194 | 58.73 |
| **Mean** | **1893.85** | 1801 | 1766 | 1754 | 1904 | 2082 | **66.91** |
| **Min** | 1162.9 | 1191 | 1098 | 949 | 920 | 1179 | 58.73 |
| **Max** | 2798.8 | 3253 | 3131 | 2761 | 3100 | 3194 | 79.88 |
| **SD** | 1236.93 | 1527 | 1209 | 1351 | 1140 | 1032 | 0.624 |
| **RMSE/Mean** | **0.1711** | 0.16 | 0.16 | 0.16 | 0.17 | 0.19 | |

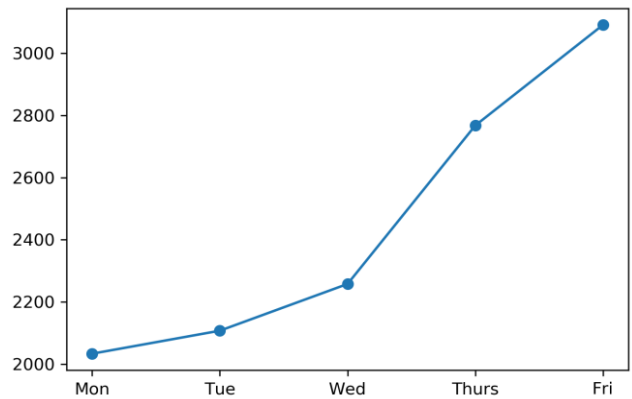

**Fig. 8.** Day-wise RMSE of LSTM#4 – multivariate encoder-decoder with previous two weeks' data as the input

Table 12 presents the performance results of the model LSTM#4 – the *multivariate encoder-decoder LSTM model*. Fig 8 depicts the weekly RMSE values of the model for round # 7 presented in Table 12. It is evident that the model is too heavy, as it took almost five times more time for the model to execute when compared to the model LSTM#3. It is also observed that the model has yielded a much higher value for the ratio of the mean RMSE to the mean *open* values when compared with other models that we discussed earlier. It is evident that the dataset of NIFTY 50 did not exhibit multivariate characteristics and univariate models were much more accurate and efficient in forecasting the future *open* values. At the same time, univariate models with one-week prior data input were very fast in their execution speed as well. The results clearly depict that while the deep learning regression models are much more accurate than the machine learning models, the univariate models with prior one-week data are among the most accurate and the fastest in execution. The univariate LSTM model with one-week data as the input turned out to be the most optimum model – both in terms of accuracy and execution time.

# 6 Conclusion

In this paper, we have presented several approaches to prediction of stock index values its movement patterns on a weekly forecast horizon using eight machine learning, and four LSTM-based deep learning regression models. Using the daily historical data of NIFTY 50 index values during the period December 29, 2014 till July 31, 2020, we constructed, optimized, and then tested the predictive models. Data pre-processing and data wrangling operations were carried on the raw data, and a set of derived variables are created for building the models. Among all the machine learning and deep learning-based regression models, the performances of the LSTM-based deep learning regression models were found to be far too superior to that of the machine-learning-based predictive models. The study has conclusively proved our conjecture that deep learning-based models have much higher capability in extracting and learning the features of a time series data than their corresponding machine learning counterparts. It also reveals the fact that multivariate analysis is not a good idea in LSTM-based regression, as univariate models are more accurate and faster in their execution. As a future scope of work, we will investigate the possibility of using *generative adversarial networks* (GANs) in time series analysis and forecasting of stock prices.